\documentclass[twocolumn]{aastex631}

\usepackage{xcolor}
\usepackage{comment}

\newcommand{\teffmath}{T_{\!\mbox{\scriptsize\it eff}}}
\newcommand{\teff}{$T_{\!\mbox{\scriptsize\it eff}}$}	
\newcommand{\logg}{log\,$g$}
\newcommand{\loggf}{log\,$g_{\!\mbox{\tiny\it F}}$}

\newcommand{\mbol}{$m\,_{\!\mbox{\tiny\scriptsize bol}}$}

\newcommand{\msun}{$M_\odot$}

\newcommand{\hii}{H\,{\sc ii}\rm}

\newcommand{\hei}{He\,{\sc i}\rm}

\newcommand{\oiii}{[O\,{\sc iii}]}

\newcommand{\feii}{Fe\,{\sc ii}\rm}

\newcommand{\eg}{{e.g.}}

\newcommand{\trgb}{\mbox{\small TRGB}}
\newcommand{\bsg}{\mbox{\small BSG}}

\newcommand{\lin}{$\,\lambda$}

\newcommand{\rtf}{$r_{25}$}

\newcommand{\re}{$r_e$}

\newcommand{\rtwothree}{R$_{23}$}

\renewcommand{\eg}{\mbox{e.g.}}

\newcommand{\fwhm}{\mbox{\small FWHM}}

\newcommand{\fglr}{\mbox{\small FGLR}}

\newcommand{\newhash}{%
	{\settoheight{\dimen0}{C}\kern-.05em \resizebox{!}{\dimen0}{\raisebox{\depth}{\#}}}}

\shorttitle{NGC~4258}
\shortauthors{Kudritzki et al.}

\begin{document}

\title{The Hubble constant anchor galaxy NGC~4258: metallicity and distance from blue supergiants}



\correspondingauthor{Rolf-Peter Kudritzki}
\email{kud@ifa.hawaii.edu}

\author{Rolf-Peter Kudritzki}
\affiliation{Institute for Astronomy, University of Hawaii, Honolulu, HI 96822, USA}
\affiliation{University Observatory Munich, Ludwig-Maximilian-Universit\"at M\"unchen, D-81679 Munich, Germany}

\author[0000-0002-9424-0501]{Miguel A.~Urbaneja}
\affil{Universit\"at Innsbruck, Institut f\"ur Astro- und Teilchenphysik, 6020 Innsbruck, Austria}

\author[0000-0002-5068-9833]{Fabio Bresolin}
\affiliation{Institute for Astronomy, University of Hawaii, Honolulu, HI 96822, USA}

\author[0000-0002-1775-4859]{Lucas M.~Macri}
\affiliation{NSF NOIRLab, Tucson, AZ 85719, USA}

\author[0000-0001-9420-6525]{Wenlong Yuan}
\affiliation{Department of Physics and Astronomy, Johns Hopkins University, Baltimore, MD 21218, USA}

\author[0000-0002-8623-1082]{Siyang Li}
\affiliation{Department of Physics and Astronomy, Johns Hopkins University, Baltimore, MD 21218, USA}

\author[0000-0002-5259-2314]{Gagandeep S.~Anand}
\affiliation{Space Telescope Science Institute, 3700 San Martin Drive, Baltimore, MD 21218, USA}

\author[0000-0002-6124-1196]{Adam G.~Riess}
\affiliation{Department of Physics and Astronomy, Johns Hopkins University, Baltimore, MD 21218, USA}
\affiliation{Space Telescope Science Institute, 3700 San Martin Drive, Baltimore, MD 21218, USA}

\begin{abstract}
A quantitative spectroscopic study of blue supergiant stars in the Hubble constant anchor galaxy NGC~4258 is presented. The non-LTE  analysis of Keck I telescope LRIS spectra yields a central logarithmic metallicity (in units of the solar value) of $[Z] = -0.05\pm0.05$ and a very shallow gradient of $-(0.09\pm0.11)$\,$r$/\rtf\ with respect to galactocentric distance in units of the isophotal radius. Good agreement with the mass-metallicity relationship of star forming galaxies based on stellar absorption line studies is found. A comparison with \hii\ region oxygen abundances obtained from the analysis of strong emission lines shows reasonable agreement when the \citet{Pettini:2004} calibration is used, while the \citet{Zaritsky:1994} calibration yields values that are 0.2 to 0.3 dex larger. These results allow to put the metallicity calibration of the Cepheid Period--Luminosity relation in this anchor galaxy on a purely stellar basis. Interstellar reddening and extinction are determined using {\it HST} and {\it JWST} photometry. Based on extinction-corrected magnitudes, combined with the stellar effective temperatures and gravities we determine, we use the Flux-weighted Gravity--Luminosity Relationship (\fglr) to estimate an independent spectroscopic distance. We obtain a distance modulus $m-M = 29.38\pm0.12$ mag, in agreement with the geometrical distance derived from the analysis of the water maser orbits in the galaxy's central circumnuclear disk. 

\end{abstract}



\section{Introduction} \label{sec:intro}

The spiral galaxy NGC~4258 plays a pivotal role in the extragalactic distance scale. Modeling of Very Long Baseline Interferometry data of water masers orbiting in its circumnuclear accretion disk  have led to a precise measurement of its geometric distance (\citealt{Herrnstein:1999, Humphreys:2013}), the more recent analysis by \citet{Reid:2019} yielding $D=7.58 \pm 0.11$\,Mpc (distance modulus $29.398 \pm 0.031$~mag). NGC~4258 thus joins the Milky Way and the Magellanic Clouds as one of the fundamental anchors of the extragalactic distance scale ladder, by providing a geometric calibration of the luminosities of Cepheid variables, Tip of the Red Giant Branch (\trgb) and J-Branch Asymptotic Giant Branch (JABG) stars, from which the Hubble constant, $H_0$, can be derived (\citealt{Riess:2022, Breuval:2024, Freedman:2024, Lee:2024}).

Our ongoing project on the quantitative spectroscopy of massive, evolved supergiant stars aims at deriving stellar metallicities and distances for a selection of nearby star-forming galaxies (\eg\ \citealt{Bresolin:2001, Bresolin:2016, Bresolin:2022, Kudritzki:2003, Kudritzki:2008, Kudritzki:2012, Kudritzki:2013,Kudritzki:2016, Urbaneja:2008, Urbaneja:2017, Urbaneja:2023, U:2009, Hosek:2014, Berger:2018, Liu:2022}). This is accomplished via relatively low resolution ($\sim$5\,\AA\ \fwhm) spectra at blue wavelengths (approximately 3800-5100\,\AA), where diagnostics of stellar temperature, gravity and metal content can be found.

One of our main objectives is the comparison between the metal content of galaxies derived from the absorption line analysis of young stars and the chemical composition derived from the emission line analysis of giant and supergiant \hii\ regions. This is a crucial endeavor, since the variety of emission line diagnostic calibrations that are available leads to large discrepancies in the nebular abundances that are derived for star-forming galaxies (\citealt{Teimoorinia:2021, Kewley:2008}). In our work we find that nebular abundances that are anchored to the measurement of electron temperatures from auroral lines such as \oiii\lin4363 are in good agreement with the stellar metallicities we derive in the same target galaxies, over more than one order of magnitude in O/H (\citealt{Bresolin:2022}). For the specific case of NGC~4258 and the distance scale problem, the \hii\ region abundance study of \citet{Bresolin:2011} showed that the abundance gradient of this galaxy is too shallow ($-0.012$~dex\,kpc$^{-1}$) to assess in a reliable way the metallicity dependence of Cepheid luminosities from the analysis of Cepheids located at different galactocentric distances. With an enlarged sample of \hii\ regions \citet{Yuan:2022} reached a similar conclusion. Nevertheless, it is still desirable to obtain a stellar-based metallicity of this galaxy, given its role as Cepheid luminosity calibrator.

Because of its special significance for extragalactic distance determinations, NGC~4258 represents the most ambitious target of our project to date. Our analysis depends on the availability of high-quality spectra (signal-to-noise ratios larger than 40 or so), which limits our work to distances $D < 8$-9\,Mpc
with the current generation of instruments and telescopes. 

In \citet{Kudritzki:2013} we demonstrated the feasibility of the quantitative stellar analysis at the distance of NGC~4258, calculating the metallicity of a single A-type supergiant. We now present the full set of observational data, derive metallicities for a sample of 12 stars and measure a spectroscopic distance to the galaxy based on the flux-weighted gravity--luminosity relationship (\fglr) technique (\citealt{Kudritzki:2003, Urbaneja:2017}). 

\section{Observations, data reduction and classification} \label{sec:observations}

\subsection{Target selection}\label{subsec:selection}
This work is the first to search for blue supergiant (\bsg) candidates in NGC~4258. For the selection of spectroscopic targets we relied on preliminary photometry from {\it Hubble Space Telescope (HST)} Advanced Camera for Surveys (ACS) fields, looking for isolated stars with $B$ magnitudes and $B-V$ color indices compatible with those expected for \bsg s\ at the distance of the galaxy. 

\begin{figure*}[ht]
	\center \includegraphics[width=0.82\textwidth]{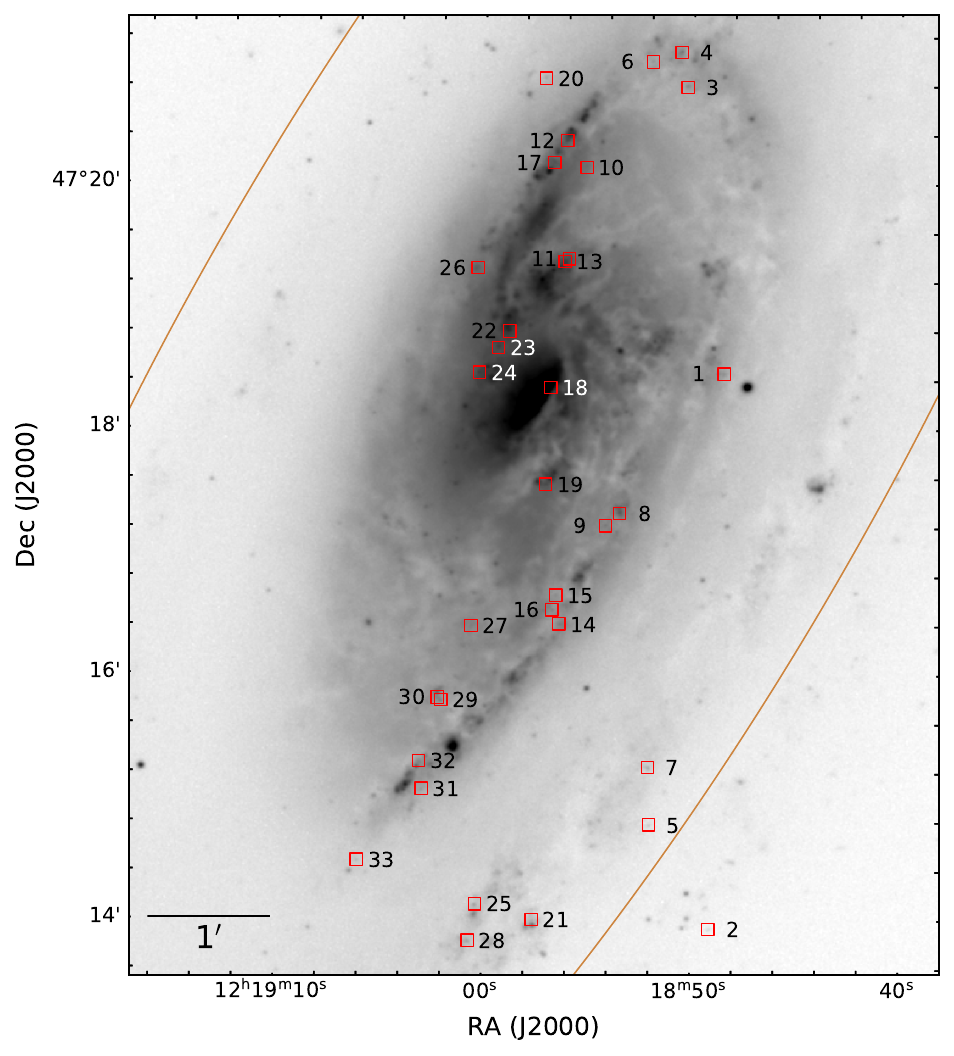}\medskip
	\caption{Identification of the spectroscopic targets in a $B$-band image of NGC~4258 taken from the NASA/IPAC Extragalactic Database (telescope: KPNO 0.9m; observers: van Zee and Barnes). The outer ring represents the projected circle having a radius equal to the isophotal radius \rtf.}\label{fig:targets}
\end{figure*}

The celestial coordinates, galactocentric distances, preliminary $B$ magnitudes and spectral types (Sec.~\ref{sec:classification}) of the final set of selected targets are reported in Table~\ref{table:1}, ordered by increasing right ascension. Figure~\ref{fig:targets} shows the spatial location of these objects.

\floattable
\begin{deluxetable}{lCCccccc}
\tabletypesize{\footnotesize}	
	\tablewidth{0pt}
	\tablecaption{Properties of potential BSG targets selected for spectroscopy. \label{table:1}}
	
	\tablehead{
		\colhead{ID}	     		&
		\colhead{R.A.}	 			&
		\colhead{Decl.}	 			&
		\colhead{$r$/\rtf}			&
		\colhead{$r$/\re}			&
		\colhead{$r$}				&
		\colhead{$B$}				&		
		\colhead{Spectral}\\[-2.5ex]		
		\colhead{}       			&
		\colhead{(J2000.0)}       	&
		\colhead{(J2000.0)}       	&
		\colhead{}					& 	
		\colhead{}					& 
		\colhead{(kpc)}				& 	
		\colhead{(mag)}            	 	& 															 											\colhead{type}	}		 	 															
  
	\colnumbers
	\startdata
	\\[-4.5ex]
    01  &       12\; 18\; 48.50 & 47\; 18\; 27.3 & 0.43 & 0.89 &  8.91 & 21.44 & composite          \\[-0.4ex] 
02  &       12\; 18\; 49.10 & 47\; 13\; 55.7 & 1.22 & 2.49 & 25.06 & 21.46 & F0                 \\[-0.4ex] 
03  &       12\; 18\; 50.32 & 47\; 20\; 47.4 & 0.31 & 0.64 &  6.41 & 21.37 & B0                 \\[-0.4ex] 
04  &       12\; 18\; 50.62 & 47\; 21\; 04.5 & 0.36 & 0.73 &  7.30 & 21.25 & B2 Ib (+WC?)       \\[-0.4ex] 
05 $\ast$ & 12\; 18\; 51.98 & 47\; 14\; 46.6 & 0.92 & 1.89 & 18.96 & 21.75 & B9                 \\[-0.4ex] 
06  &       12\; 18\; 52.00 & 47\; 20\; 59.8 & 0.37 & 0.75 &  7.49 & 21.17 & F0                 \\[-0.4ex] 
07  &       12\; 18\; 52.05 & 47\; 15\; 14.6 & 0.83 & 1.70 & 17.05 & 21.06 & emission line star \\[-0.4ex] 
08 $\ast$ & 12\; 18\; 53.47 & 47\; 17\; 18.9 & 0.37 & 0.75 &  7.58 & 21.29 & B6/B7 ?composite?  \\[-0.4ex] 
10  &       12\; 18\; 55.16 & 47\; 20\; 08.1 & 0.29 & 0.59 &  5.91 & 22.03 & late F/G           \\[-0.4ex] 
11  &       12\; 18\; 55.97 & 47\; 19\; 23.2 & 0.17 & 0.35 &  3.51 & 21.65 & B5 Ib              \\[-0.4ex] 
12  &       12\; 18\; 56.09 & 47\; 20\; 21.0 & 0.36 & 0.74 &  7.42 & 20.95 & B1                 \\[-0.4ex] 
13  &       12\; 18\; 56.18 & 47\; 19\; 21.9 & 0.17 & 0.35 &  3.55 & 21.81 & composite          \\[-0.4ex] 
14 $\ast$ & 12\; 18\; 56.36 & 47\; 16\; 24.6 & 0.41 & 0.83 &  8.38 & 21.74 & A1                 \\[-0.4ex] 
15  &       12\; 18\; 56.52 & 47\; 16\; 38.7 & 0.36 & 0.73 &  7.29 & 21.98 & composite          \\[-0.4ex] 
16 $\ast$ & 12\; 18\; 56.70 & 47\; 16\; 31.5 & 0.37 & 0.76 &  7.62 & 21.42 & B7                 \\[-0.4ex] 
17  &       12\; 18\; 56.71 & 47\; 20\; 10.3 & 0.35 & 0.71 &  7.18 & 21.95 & composite (WR)     \\[-0.4ex] 
18  &       12\; 18\; 56.83 & 47\; 18\; 20.2 & 0.02 & 0.05 &  0.48 & 20.19 & composite          \\[-0.4ex] 
19  &       12\; 18\; 57.04 & 47\; 17\; 33.0 & 0.16 & 0.32 &  3.18 & 20.52 & composite (WR)     \\[-0.4ex] 
20 $\ast$ & 12\; 18\; 57.14 & 47\; 20\; 51.5 & 0.50 & 1.03 & 10.30 & 21.34 & A2                 \\[-0.4ex] 
21 $\ast$ & 12\; 18\; 57.59 & 47\; 13\; 59.9 & 0.83 & 1.70 & 17.07 & 22.01 & A1 Ib              \\[-0.4ex] 
22  &       12\; 18\; 58.81 & 47\; 18\; 47.7 & 0.17 & 0.34 &  3.46 & 21.34 & composite          \\[-0.4ex] 
23 $\ast$ & 12\; 18\; 59.35 & 47\; 18\; 39.6 & 0.17 & 0.35 &  3.47 & 21.14 & A5                 \\[-0.4ex] 
24  &       12\; 19\; 00.25 & 47\; 18\; 27.5 & 0.18 & 0.36 &  3.66 & 20.78 & A3                 \\[-0.4ex] 
25  &       12\; 19\; 00.32 & 47\; 14\; 07.5 & 0.70 & 1.43 & 14.41 & 21.26 & emission line star \\[-0.4ex] 
26 $\ast$ & 12\; 19\; 00.36 & 47\; 19\; 18.8 & 0.34 & 0.70 &  6.99 & 21.01 & A3                 \\[-0.4ex] 
27  &       12\; 19\; 00.58 & 47\; 16\; 23.5 & 0.26 & 0.53 &  5.30 & 21.56 & emission line star \\[-0.4ex] 
28  &       12\; 19\; 00.66 & 47\; 13\; 49.7 & 0.75 & 1.53 & 15.35 & 21.97 & B3 Ib              \\[-0.4ex] 
29 $\ast$ & 12\; 19\; 02.01 & 47\; 15\; 47.2 & 0.33 & 0.68 &  6.83 & 20.43 & A3/A5              \\[-0.4ex] 
30  &       12\; 19\; 02.18 & 47\; 15\; 48.5 & 0.33 & 0.66 &  6.68 & 21.41 & composite (WR)     \\[-0.4ex] 
31 $\ast$ & 12\; 19\; 02.92 & 47\; 15\; 03.8 & 0.44 & 0.90 &  9.06 & 22.05 & B7/B8              \\[-0.4ex] 
32 $\ast$ & 12\; 19\; 03.06 & 47\; 15\; 17.3 & 0.40 & 0.81 &  8.16 & 21.33 & B8 Ib              \\[-0.4ex] 
33 $\ast$ & 12\; 19\; 06.00 & 47\; 14\; 28.9 & 0.48 & 0.98 &  9.84 & 21.75 & B9                 \\[-0.4ex] 
    \\[-2.5ex]
	\enddata
	\tablecomments{Normalized galactocentric distances adopt the following disk geometry: i\,=\,72~deg, PA\,=\,150~deg (\citealt{van-Albada:1980}), \rtf\,=\,558 arcsec (\citealt{de-Vaucouleurs:1991}) and \re\,=\,274 arcsec (\citealt{Watkins:2016}). Unless the luminosity class is specified, the spectra are consistent with the Ia class. Stars identified with the $\ast$ symbol in column (1) are those analyzed in Sec.~\ref{sec:quantitative}. The $B$ magnitudes are the preliminary values used for target selection. More precise magnitudes are given in Table \ref{table:2}. }
\end{deluxetable}


\subsection{Spectroscopic observations and classification\label{sec:classification}}
Our spectroscopic data were obtained with the Low Resolution Imaging Spectrometer (LRIS, \citealt{Oke:1995}) at the Keck~I telescope, during the course
of nine full nights, distributed over four distinct observing campaigns in 2011, 2012, 2013 and 2015. The target list used to define the multi-object mask 
employed for the observations was revised during this time-span, which resulted in different total exposure times for different targets. The number of slits used during each run varied between 25 and 27, with several slits covering \hii\ regions and stellar cluster candidates. These will be excluded from the following analysis, so that the final number of stellar targets considered in the remainder of this work is 33.  

While both blue and red channel multi-object data were gathered, our paper only presents results based on the blue channel data, obtained with the 600/4000 grism, and covering the approximate wavelength range 3300--5600~\AA. During each observing night we obtained multiple exposure times, of 1800\,s or 2700\,s in duration, as long as the airmass of NGC~4258 was smaller than approximately 1.8. A slit width of 1\farcs2 yielded a \fwhm\ spectral resolution of $\sim$5\,\AA. We estimated the seeing from the spatial profile of the stellar spectra to vary between 0\farcs8 and 1\farcs6, with a mean value of 1\farcs2.
Due to the evolving target list, the total exposure time per target varied between 6.0\,h and 42.8\,h, with a median of 19.6\,h.

The spectral data were reduced with {\sc iraf}\footnote{IRAF is distributed by the Community Science and Data Center at NSF NOIRLab, which is managed by the Association of Universities for Research in Astronomy (AURA) under a cooperative agreement with the U.S. National Science Foundation.}, by carrying out the bias subtraction, flat field correction and wavelength calibration. The spectral extractions for individual exposures were registered with the aid of sky lines and subsequently combined. The final steps consisted in the flux rectification with low-order polynomials and a Doppler correction that utilized stellar absorption lines to yield spectra in the rest frame.
The average signal-to-noise ratio of the co-added spectra ranges between 30 and 120 (median: 64).

The spectral types of the bona fide stars included in Table~\ref{table:1} were estimated following the MK criteria, with the aid of the monograph by \citet{Gray:2009}. The resulting classification is listed in Column (8) of the table, where we have used the term 'composite' in those instances where we can identify features originating from more than one spectral type. Four cases involve stellar absorption features appearing together with broad Wolf-Rayet (WC type) emission at 4650\,\AA.
Comparison with spectral standards in digital format (\citealt{Gray:2014}) provides a Ia luminosity classification for most targets, with the exception of a minority of Ib stars. 

The spectra of three targets, marked as emission line stars in Table~\ref{table:1}, display Balmer and \hei\ lines in emission, in some cases with evidence of P~Cygni profiles, together with a large number of  \feii\ emission lines, as seen, for example, in the spectra of the Galactic iron stars HD~87643 and GC~Car displayed by \citet{Walborn:2000a}. The spectrum of star 07 resembles the spectrum of the Luminous Blue Variable AG~Carinae near maximum (\citealt{Wolf:1982, Stahl:2001}).

\subsection{Photometry \label{sec:phot}}

In addition to the spectroscopy, used for the determination of stellar temperatures, gravities and metallicities, we also need multi-band photometry to constrain interstellar reddening, extinction and luminosities. We analyzed all available {\it HST} and {\it JWST} images of NGC~4258 containing the objects from Table~\ref{table:1}, obtained with ACS, WFC3 and NIRCam in 11 different filters spanning $0.4-3.5\mu$m, as part of programs HST-GO-10399 (PI: Greenhill; ACS {F606W}), HST-GO-11570 (PI: Riess; ACS {F435W, F555W, F814W}) and WFC3/IR {F160W}), HST-GO-13445 (PI: Bloom; WFC3/IR F160W); HST-GO-16198 (PI: Riess; WFC3/IR F160W) and JWST-GO-2875 (PI: Riess; NIRCam {F070W, F090W, F115W, F150W, F277W, F356W}). The images were downloaded from the Mikulski Archive for Space Telescopes (MAST) at the Space Telescope Science Institute. The specific observations used in this work can be accessed via this \dataset[DOI]{https://dx.doi.org/10.17909/jysd-eq19}, and were analyzed as described below.
\clearpage

\subsubsection{HST photometry}

{\it HST}/ACS images covering a given section of the galaxy in a given filter were combined into mosaics using {\sc DrizzlePac}  at the native CCD scale of $0\farcs 05/$pix. The ACS {F606W} images of interest contain 10 targets from Table~\ref{table:1} and were obtained as 12 cosmic-ray split pairs of parallel observations. Each pair was drizzled into a combined image. The ACS {F435W, F555W, F814W} images were obtained as part of a galaxy-wide survey, with single-image observations in {F435W} and {F814W} and dithered triples in {F555W}. We drizzled various images in a given band into four mosaics, such that each object would appear in at least one region covered by multiple exposures (and therefore mostly free from cosmic rays).

Aperture photometry will yield the most precise and accurate magnitude measurements for our targets at optical wavelengths, since they are bright and fairly isolated, and fainter nearby sources can be removed. We used {\sc DAOPHOT} \citep{Stetson:1987} to detect all sources in each mosaic with a threshold of $3\sigma$, and {\sc ALLSTAR} \citep{Stetson:1994} to carry out point-spread function (PSF) photometry. The PSFs were simulated using {\sc TinyTim} \citep{Krist:1993}. We subtracted all detected sources from each mosaic, leaving only the BSGs and $\sim$20-40 bright and isolated stars depending on the image and filter. The latter were used to derive curves of growth by carrying out aperture photometry at ten radii spanning $3-5$~pix ($0\farcs15-0\farcs25$). We then carried out aperture photometry of the BSGs at radii of 3, 4 and 5 pixels, correcting the first two to the larger aperture using the growth curves. The derived magnitudes were consistent to better than 0.01~mag in most cases, except for some objects embedded in nebulosities. In those cases we only used the measurement from the smallest aperture, suitably corrected via the growth curves.

Many of the targets appeared in multiple mosaics, allowing us to check the consistency of the photometry. The error-weighted mean differences were less than 0.01~mag, with scatter of 0.016-0.035~mag depending on the filter. Based on this comparison, we assigned minimum photometric uncertainties of 0.02, 0.015, 0.025 and 0.015~mag in {F435W, F555W, F606W} and {F814W}, respectively. Lastly, we calibrated our measurements into the ACS ``infinite aperture'' VEGAMag system using the encircled energy values derived by \citet{Bohlin:2016} and the zeropoints provided by the STScI calculator\footnote{\url{https://acszeropoints.stsci.edu/}}.

As a next step, we analyzed the WFC3/IR {F160W} images containing our targets. Due to the lower angular resolution of the telescope at these wavelengths and the coarser pixel sampling of this detector compared to  ACS, the stellar field is considerably more crowded than in the optical bands. In order to improve the photometric accuracy, we first derived the object positions in the better-resolved {F814W} images through PSF fitting, then mapped their coordinates to the near-infrared images through frame-to-frame transformations obtained from all sources in common. With the known positions in the {F160W} images, we are able to fit only one parameter (flux) per target instead of three (flux, x position, y position) to the PSF model, and thus obtain improved flux measurements.

Additionally, we computed possible photometric biases originating from crowding through extensive artificial star tests. We randomly placed 1,000 artificial stars of known brightness in the vicinity of each object at distances spanning $0.4-1.6\arcsec$ and recovered their magnitudes following the same photometry procedures described above. We calculated the bias correction as the unweighted mean of the differences between the input and output magnitudes of artificial stars, applying an iterative $2.7\sigma$ clipping. The procedure is similar to those carried out by \citet{Riess:2009}, \citet{Riess:2011}, \citet{Riess:2016} and \citet{Riess:2022}. The uncertainty in the mean of the recovered artificial star magnitudes was added in quadrature to the photometric uncertainty of each object reported by {\sc ALLSTAR}. The median correction was 0.019~mag, a small value thanks to the high luminosities of the BSGs.

Table \ref{table:2} presents the mean magnitudes and uncertainties for all targets present in the {\it HST} data.

\subsubsection{{\it JWST} photometry}

The {\it JWST} photometry was carried out following the same procedures and quality cuts listed in detail in \citet{Riess:2024}, \citet{Anand:2024} and \citet{Li:2024}. We performed PSF photometry using the {\it JWST}/NIRCAM module \citep{Weisz:2024} of the {\sc DOLPHOT} software package \citep{Dolphin:2000,Dolphin:2016} and calibrated the magnitudes using Vega-Vega zeropoints\footnote{\url{https://jwst-docs.stsci.edu/jwst-near-infrared-camera/nircam-performance/nircam-absolute-flux-calibration-and-zeropoints}} using the ``jwst\_1126.pmap" delivered on 2023 Sep 14.

The photometry was performed using the ``warmstart'' option, where the source list derived from only the short-wavelength (SW) channels was used to perform photometry on the combined dataset to improve PSF subtraction \citep{Riess:2023}. The adopted photometry consists of the SW photometry from the SW-only run, and the long-wavelength (LW) photometry from the combined run.

Lastly, we carried out artificial star simulations as described in the preceding subsection to correct for photometric biases due to crowding, placing 100 artificial stars in the vicinity of each object. The procedure is the same as described in \citet{Riess:2023}. The median corrections ranged from 0.005~mag in F070W to 0.027~mag in F356W. As for the {\it HST} crowding corrections, they are small because the BSG are extremely bright. The corrected magnitudes are provided in Table \ref{table:3}.


\begin{deluxetable*}{lCCccc}
	\tabletypesize{\footnotesize}	
	\tablecolumns{6}
	\tablewidth{0pt}
	\tablecaption{{\it HST} photometry.\label{table:2}}
	
	\tablehead{
		\colhead{ID}	     		&
		\colhead{m$_{F435W}$}	 			&
		\colhead{m$_{F555W}$}	 			&
        \colhead{m$_{F606W}$}	 			&
		\colhead{m$_{F814W}$}			&
		\colhead{m$_{F160W}$}\\[-2.ex]
		\colhead{}       			&
		\colhead{(mag)}       	&
		\colhead{(mag)}       	&
        \colhead{(mag)}       	&
		\colhead{(mag)}				& 														
		\colhead{(mag)}  } 	
	
	\colnumbers
	\startdata
	\\[-4.5ex]
	03 & $21.41\pm0.02$ & $21.44\pm0.02$ & \nodata          & $21.54\pm0.02$ & $21.86\pm0.03$ \\[-0.4ex]
04 & $21.28\pm0.02$ & $21.28\pm0.02$ & \nodata          & $21.23\pm0.02$ & $20.79\pm0.05$ \\[-0.4ex]
05 & $21.90\pm0.02$ & $21.87\pm0.02$ & $21.88\pm0.03$ & $21.95\pm0.03$ & $21.93\pm0.05$ \\[-0.4ex]
06 & $21.45\pm0.02$ & $21.29\pm0.02$ & \nodata          & $21.11\pm0.02$ & $20.21\pm0.03$ \\[-0.4ex]
08 & $21.53\pm0.02$ & $21.39\pm0.02$ & $21.37\pm0.03$ & $21.16\pm0.02$ & $21.14\pm0.10$ \\[-0.4ex]
10 & $22.27\pm0.02$ & $22.17\pm0.02$ & \nodata          & $21.93\pm0.02$ & $21.85\pm0.13$ \\[-0.4ex]
11 & $21.70\pm0.02$ & $21.66\pm0.02$ & \nodata          & $21.55\pm0.02$ & $21.61\pm0.05$ \\[-0.4ex]
12 & $21.25\pm0.02$ & $20.90\pm0.02$ & \nodata          & $19.84\pm0.02$ & $18.15\pm0.05$ \\[-0.4ex]
13 & $21.89\pm0.02$ & $21.88\pm0.02$ & \nodata          & $21.84\pm0.02$ & $21.79\pm0.09$ \\[-0.4ex]
14 & $21.93\pm0.02$ & $21.81\pm0.02$ & $21.72\pm0.03$ & $21.52\pm0.02$ & $21.06\pm0.11$ \\[-0.4ex]
16 & $21.60\pm0.02$ & $21.50\pm0.02$ & $21.48\pm0.03$ & $21.44\pm0.02$ & $21.41\pm0.05$ \\[-0.4ex]
17 & $22.07\pm0.03$ & $22.03\pm0.03$ & \nodata          & $21.95\pm0.04$ & $22.40\pm0.06$ \\[-0.4ex]
20 & $21.65\pm0.02$ & $21.25\pm0.02$ & \nodata          & $20.72\pm0.03$ & $20.16\pm0.07$ \\[-0.4ex]
21 & $22.13\pm0.02$ & $22.04\pm0.02$ & $22.20\pm0.03$ & $21.85\pm0.02$ & $21.72\pm0.13$ \\[-0.4ex]
23 & $21.59\pm0.02$ & $21.04\pm0.02$ & $20.85\pm0.03$ & $20.39\pm0.02$ & $20.08\pm0.07$ \\[-0.4ex]
26 & $21.49\pm0.02$ & $21.14\pm0.02$ & $20.96\pm0.03$ & $20.70\pm0.02$ & $20.45\pm0.07$ \\[-0.4ex]
29 & $20.85\pm0.02$ & $20.58\pm0.02$ & $20.48\pm0.04$ & $20.10\pm0.02$ & $19.95\pm0.05$ \\[-0.4ex]
31 & $22.03\pm0.02$ & $22.09\pm0.02$ & $22.11\pm0.03$ & $22.04\pm0.02$ & $22.32\pm0.05$ \\[-0.4ex]
32 & $21.42\pm0.02$ & $21.37\pm0.02$ & $21.36\pm0.03$ & $21.27\pm0.02$ & $21.55\pm0.04$ \\[-0.4ex]
33 & $21.88\pm0.03$ & $21.83\pm0.02$ & \nodata          & $21.71\pm0.02$ & $21.59\pm0.09$ \\[-0.4ex]

	\\[-2.5ex]
	\enddata
\end{deluxetable*}
\begin{deluxetable*}{lCCcccc}
	\tabletypesize{\footnotesize}	
	\tablecolumns{7}
	\tablewidth{0pt}
	\tablecaption{JWST photometry.\label{table:3}}
	
	\tablehead{
		\colhead{ID}	     		&
		\colhead{$m_{F070W}$}	 			&
		\colhead{$m_{F090W}$}	 			&
		\colhead{$m_{F115W}$}			&
		\colhead{$m_{F150W}$}			&
        \colhead{$m_{F277W}$ }       &
        \colhead{$m_{F356W}$}   \\[-2.ex]
        \colhead{ }      &     
		\colhead{(mag)}       			&
		\colhead{(mag)}       	&
		\colhead{(mag)}       	&
		\colhead{(mag)}					& 	
		\colhead{(mag)}                & 
        \colhead{(mag)}} 	
	
	\colnumbers
	\startdata
	\\[-4.5ex]
	01 & 21.51$\pm0.01$ & 21.50$\pm0.01$ & 21.51$\pm0.01$ & 21.46$\pm0.01$ & 21.34$\pm0.01$ & 21.30$\pm0.01$ \\[-0.4ex]
08 & 21.32$\pm0.01$ & 21.09$\pm0.01$ & 21.05$\pm0.01$ & 20.99$\pm0.01$ & 20.85$\pm0.01$ & 20.80$\pm0.01$ \\[-0.4ex]
09 & 22.67$\pm0.01$ & 21.87$\pm0.01$ & 21.20$\pm0.01$ & 20.52$\pm0.01$ & 20.06$\pm0.01$ & 19.77$\pm0.01$ \\[-0.4ex]
11 & 21.63$\pm0.01$ & 21.60$\pm0.01$ & 21.65$\pm0.01$ & 21.67$\pm0.01$ & 21.64$\pm0.01$ & 21.86$\pm0.03$ \\[-0.4ex]
13 & 21.72$\pm0.01$ & 21.65$\pm0.01$ & 21.68$\pm0.01$ & 21.66$\pm0.01$ & 21.50$\pm0.02$ & 21.50$\pm0.02$ \\[-0.4ex]
14 & 21.60$\pm0.01$ & 21.51$\pm0.01$ & 21.42$\pm0.01$ & 21.30$\pm0.01$ & 21.19$\pm0.01$ & 21.13$\pm0.01$ \\[-0.4ex]
16 & 21.50$\pm0.01$ & 21.56$\pm0.01$ & 21.53$\pm0.01$ & 21.50$\pm0.01$ & 21.49$\pm0.01$ & 21.54$\pm0.02$ \\[-0.4ex]
18 & 20.44$\pm0.01$ & 20.36$\pm0.01$ & 20.37$\pm0.01$ & 20.33$\pm0.01$ & 20.09$\pm0.02$ & 19.78$\pm0.02$ \\[-0.4ex]
22 & 21.50$\pm0.01$ & 20.73$\pm0.01$ & 20.01$\pm0.01$ & 19.26$\pm0.01$ & 18.55$\pm0.01$ & 18.26$\pm0.01$ \\[-0.4ex]
23 & 20.61$\pm0.01$ & 20.26$\pm0.01$ & 20.11$\pm0.01$ & 19.97$\pm0.01$ & 19.70$\pm0.01$ & 19.61$\pm0.01$ \\[-0.4ex]
24 & 20.62$\pm0.01$ & 20.37$\pm0.01$ & 20.30$\pm0.01$ & 20.22$\pm0.01$ & 20.03$\pm0.01$ & 20.01$\pm0.01$ \\[-0.4ex]
27 & 20.43$\pm0.01$ & 20.31$\pm0.01$ & 20.11$\pm0.01$ & 19.86$\pm0.01$ & 19.33$\pm0.01$ & 19.11$\pm0.01$ \\[-0.4ex]
29 & 20.35$\pm0.01$ & 20.11$\pm0.01$ & 20.01$\pm0.01$ & 19.86$\pm0.01$ & 19.64$\pm0.01$ & 19.58$\pm0.01$ \\[-0.4ex]
30 & 21.44$\pm0.01$ & 21.39$\pm0.01$ & 21.40$\pm0.01$ & 21.34$\pm0.01$ & 21.21$\pm0.01$ & 21.18$\pm0.01$ \\[-0.4ex]
31 & 22.13$\pm0.01$ & 22.17$\pm0.01$ & 22.20$\pm0.01$ & 22.17$\pm0.01$ & 22.12$\pm0.01$ & 22.10$\pm0.01$ \\[-0.4ex]
32 & 21.45$\pm0.01$ & 21.37$\pm0.01$ & 21.36$\pm0.01$ & 21.31$\pm0.01$ & 21.23$\pm0.01$ & 21.20$\pm0.01$ \\[-0.4ex]

	\\[-2.5ex]
	\enddata
\end{deluxetable*}

\begin{figure*}[t]
	\center \includegraphics[width=0.93\textwidth]{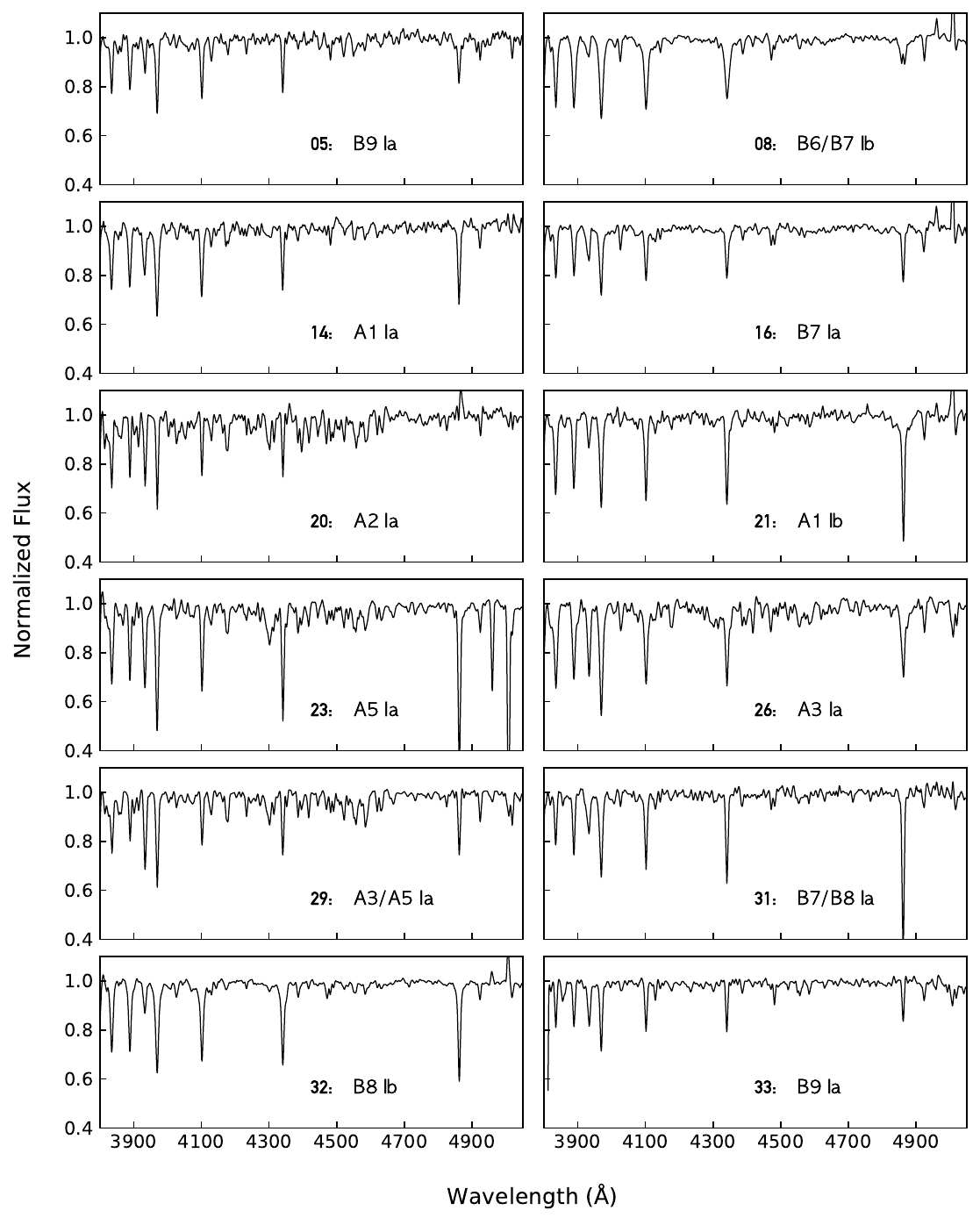}\medskip
	\caption{Spectra of the 12 targets analyzed quantitatively.}\label{fig:spectra}
\end{figure*}
\vspace*{-60pt}

\section{Quantitative spectral analysis} \label{sec:quantitative}

Table~\ref{table:1} contains 21 targets with well defined single object absorption line spectral types, which are, in principle, suitable for a detailed spectroscopic analysis. However, six of these targets (ID 03, 11, 12, 13, 24, 28) turned out to have spectra where the Balmer lines were heavily contaminated with \hii\ region hydrogen line emission. Since the Balmer lines are crucial to constrain stellar gravities, these targets could not be used for an analysis. In addition, three targets (ID 02, 06 and 10) have the relatively late spectral type F and turned out to have effective temperatures that are too low for our analysis technique (described below) to be applied. This leaves us with 12 targets for the spectroscopic analysis. These targets are indicated by asterisks in Table~\ref{table:1} and their spectra are shown in Figure~\ref{fig:spectra}.

\begin{deluxetable*}{lCCcccccc}
	\tabletypesize{\footnotesize}	
	\tablecolumns{9}
	\tablewidth{0pt}
	\tablecaption{Spectroscopic stellar parameters and reddening, extinction and apparent bolometric magnitude.\label{table:4}}
	
	\tablehead{
		\colhead{ID}	     		&
		\colhead{\teff}	 			&
		\colhead{\logg}	 			&
		\colhead{$[Z]$}			&
		\colhead{\loggf}			&
        \colhead{$BC_{F555W}$}  &
        \colhead{$E_{F435W-F555W}$}  &
        \colhead{$R_{F555W}$}        &
        \colhead{\mbol}   \\[-2.ex]
		\colhead{}       			&
		\colhead{(K)}       	&
		\colhead{(cgs)}       	&
		\colhead{(dex)}					& 	
		\colhead{(cgs)}					& 
		\colhead{(mag)}                & 
        \colhead{(mag)}          &
        \colhead{}       			&
        \colhead{(mag)}} 	
	
	\colnumbers
	\startdata
	\\[-4.5ex]
	05 &  9900$\pm200$ & 1.38$\pm0.05$  & $-0.15\pm0.10$ & $1.40\pm0.06$ & $-0.24\pm0.04$ & $0.00\pm0.01$ &      \nodata        & $21.62\pm0.07$ \\[-0.4ex] 
08 & 13500$\pm100$ & 2.10$\pm0.05$  & $-0.12\pm0.05$ & $1.58\pm0.05$ & $-0.87\pm0.02$ & $0.17\pm0.03$ & $5.54\pm0.70$ & $19.56\pm0.09$  \\[-0.4ex] 
14 &  9300$\pm200$ & 1.27$\pm0.05$  & $-0.14\pm0.07$ & $1.40\pm0.18$ & $-0.14\pm0.05$ & $0.10\pm0.01$ & $5.90\pm0.70$ & $21.08\pm0.09$ \\[-0.4ex] 
16 & 12750$\pm300$ & 1.77$\pm0.05$  & $+0.02\pm0.08$ & $1.35\pm0.05$ & $-0.76\pm0.06$ & $0.11\pm0.02$ & $2.41\pm0.60$ & $20.48\pm0.07$  \\[-0.4ex] 
20 &  8150$\pm75 $ & 0.79$\pm0.05$  & $+0.01\pm0.05$ & $1.15\pm0.09$ & $+0.06\pm0.05$ & $0.27\pm0.03$ & $3.40\pm0.70$ & $20.37\pm0.21$  \\[-0.4ex] 
21 &  9150$\pm350$ & 1.41$\pm0.05$  & $-0.30\pm0.12$ & $1.56\pm0.11$ & $-0.10\pm0.07$ & $0.08\pm0.03$ & $4.27\pm2.50$ & $21.60\pm0.14$ \\[-0.4ex] 
23 &  8100$\pm50 $ & 0.83$\pm0.05$  & $-0.07\pm0.04$ & $1.20\pm0.08$ & $+0.08\pm0.03$ & $0.34\pm0.02$ & $3.25\pm0.25$ & $20.02\pm0.06$  \\[-0.4ex] 
26 &  7985$\pm25 $ & 0.75$\pm0.05$  & $-0.26\pm0.04$ & $1.14\pm0.07$ & $+0.07\pm0.02$ & $0.24\pm0.03$ & $2.07\pm0.70$ & $20.71\pm0.16$ \\[-0.4ex] 
29 &  8425$\pm25 $ & 0.87$\pm0.05$  & $+0.07\pm0.04$ & $1.17\pm0.06$ & $-0.01\pm0.03$ & $0.14\pm0.02$ & $5.11\pm0.70$ & $19.87\pm0.06$ \\[-0.4ex] 
31 & 11400$\pm200$ & 1.68$\pm0.05$  & $-0.10\pm0.07$ & $1.45\pm0.05$ & $-0.51\pm0.04$ & $0.02\pm0.01$ & $6.09\pm1.10$ & $21.47\pm0.07$  \\[-0.4ex] 
32 & 10950$\pm200$ & 1.74$\pm0.05$  & $-0.10\pm0.08$ & $1.58\pm0.06$ & $-0.40\pm0.04$  & $0.04\pm0.01$ & $6.83\pm1.50$ & $20.72\pm0.09$ \\[-0.4ex] 
33 & 10300$\pm100$ & 1.41$\pm0.05$  & $-0.05\pm0.06$ & $1.36\pm0.05$ & $-0.32\pm0.03$  & $0.06\pm0.03$ & $5.97\pm2.50$ & $21.18\pm0.12$ \\[-0.4ex] 
	\\[-2.5ex]
	\enddata
	\tablecomments{The \logg\ uncertainty is for a fixed \teff\ value.}
\end{deluxetable*}
\vspace*{-24pt}
In our analysis we proceed in the usual way as described in the publications mentioned in the introduction (see \citealt{Bresolin:2022} and \citealt{Urbaneja:2023} as the most recent examples). In short, we normalize the observed spectra to a continuum which is constant as a function of wavelength and compare with synthetic spectra obtained from a comprehensive grid of model atmospheres with extensive non-LTE line formation calculations. Details of the calculations and model grid are given in \citet{Przybilla:2006} and \citet{Kudritzki:2008, Kudritzki:2012}. We use a fit of the hydrogen Balmer line profiles to constrain stellar gravity (\logg) as a function of effective temperature (\teff) and then moving along the gravity-temperature relationship and comparing metal line-dominated spectral windows we calculate $\chi^2$ values as function of \teff\ and metallicity $[Z]$ (defined as $[Z] = \log Z/Z_{\odot}$, where $Z_{\odot}$ is the solar metallicity corresponding to a metal mass fraction of 0.014, see \citealt{Asplund:2009}). The minimum $\chi^2$ and surrounding isocontours $\Delta \chi^2$ yield \teff\ and $[Z]$ with the corresponding uncertainties.

\begin{figure}[t]
    \centering
    \includegraphics[width=0.95\linewidth]{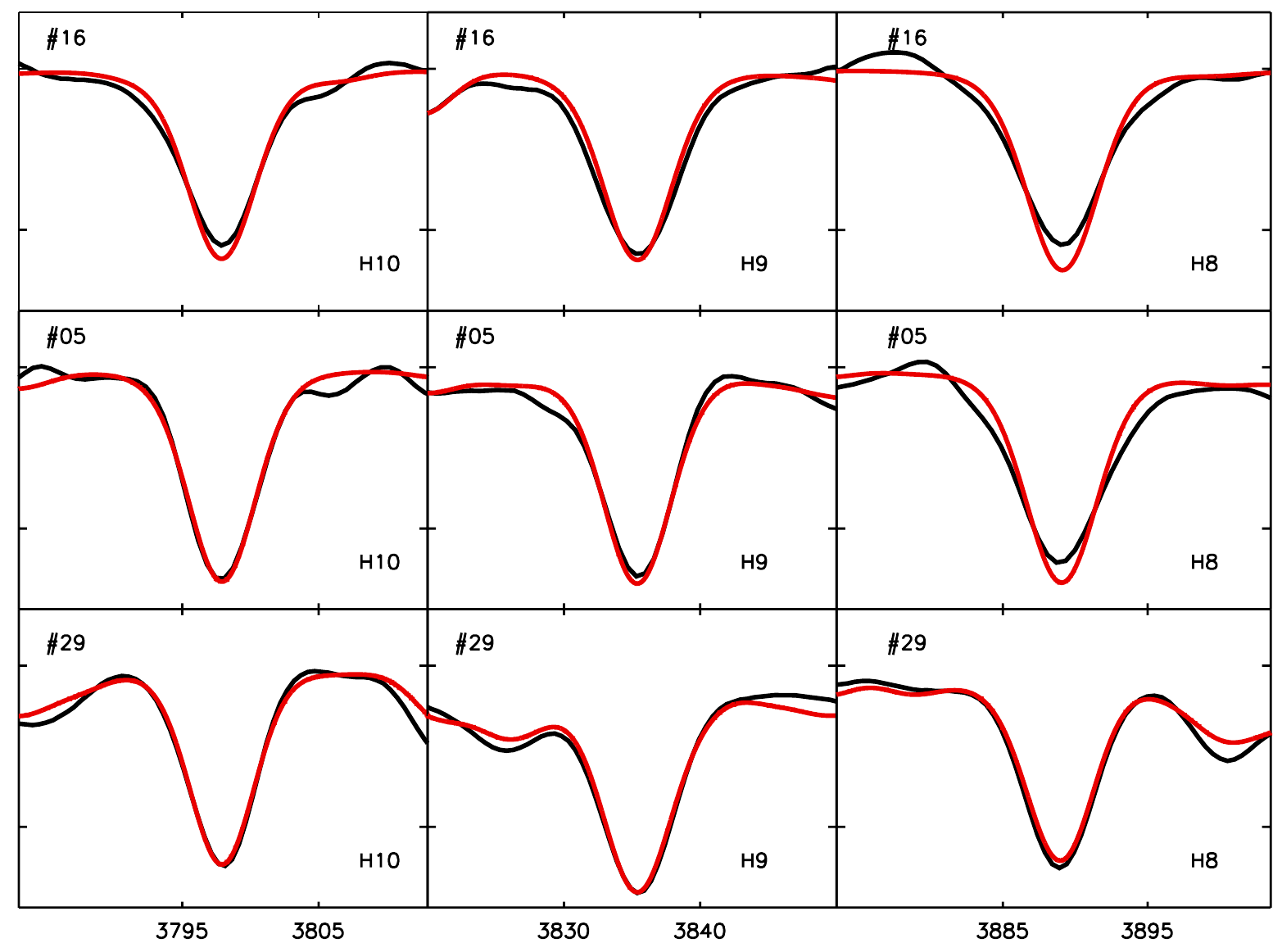}
    \caption{Fit of the higher Balmer lines H$_8$, H$_9$, and H$_{10}$ with the final model of targets ID 16 (upper row), 5 (middle row), and 29 (lower row). The tickmarks on the y-axis correspond to a value of 0.2 in units of the continuum.}
    \label{fig:Balmerfit}
\end{figure}

\begin{figure}[ht]
    \centering
    \includegraphics[width=0.445\textwidth]{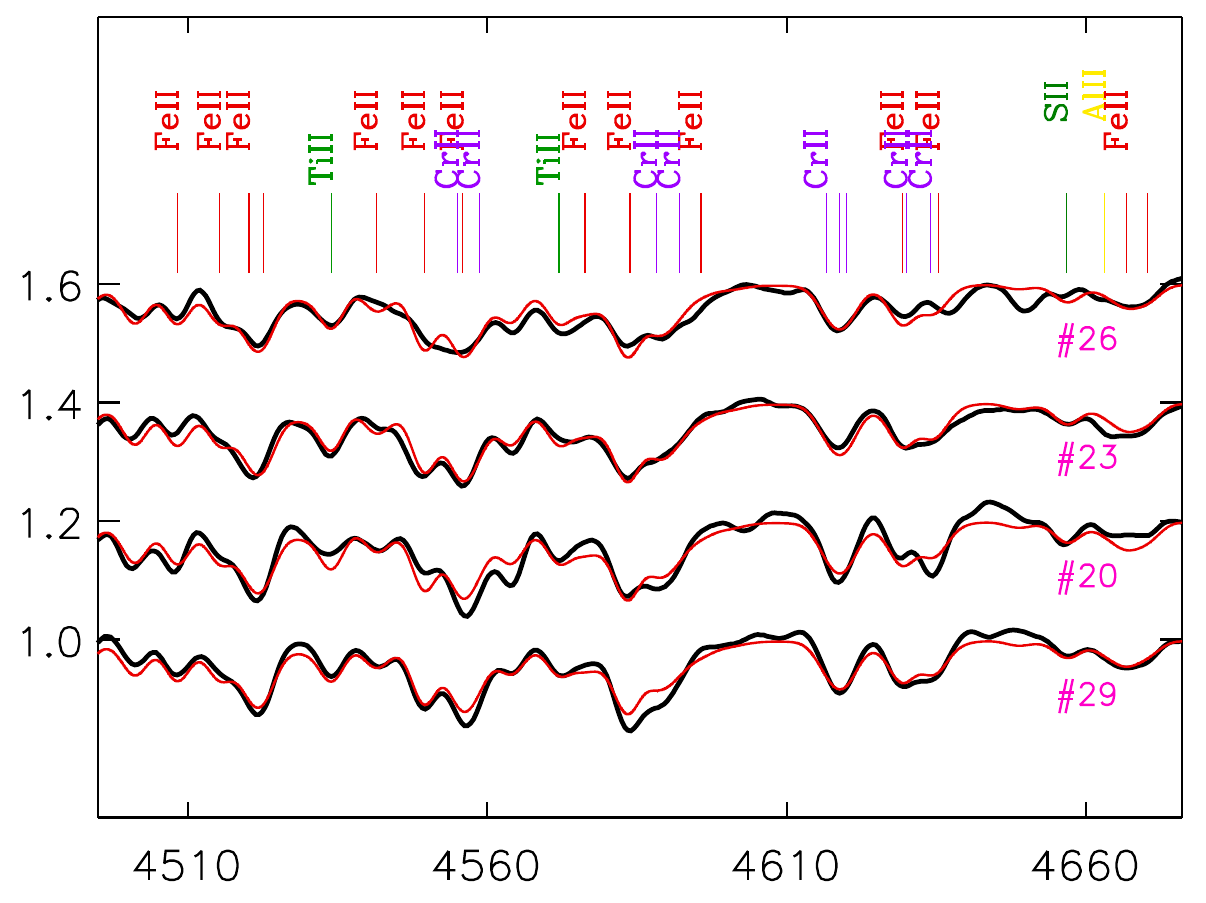}
    \includegraphics[width=0.445\textwidth]{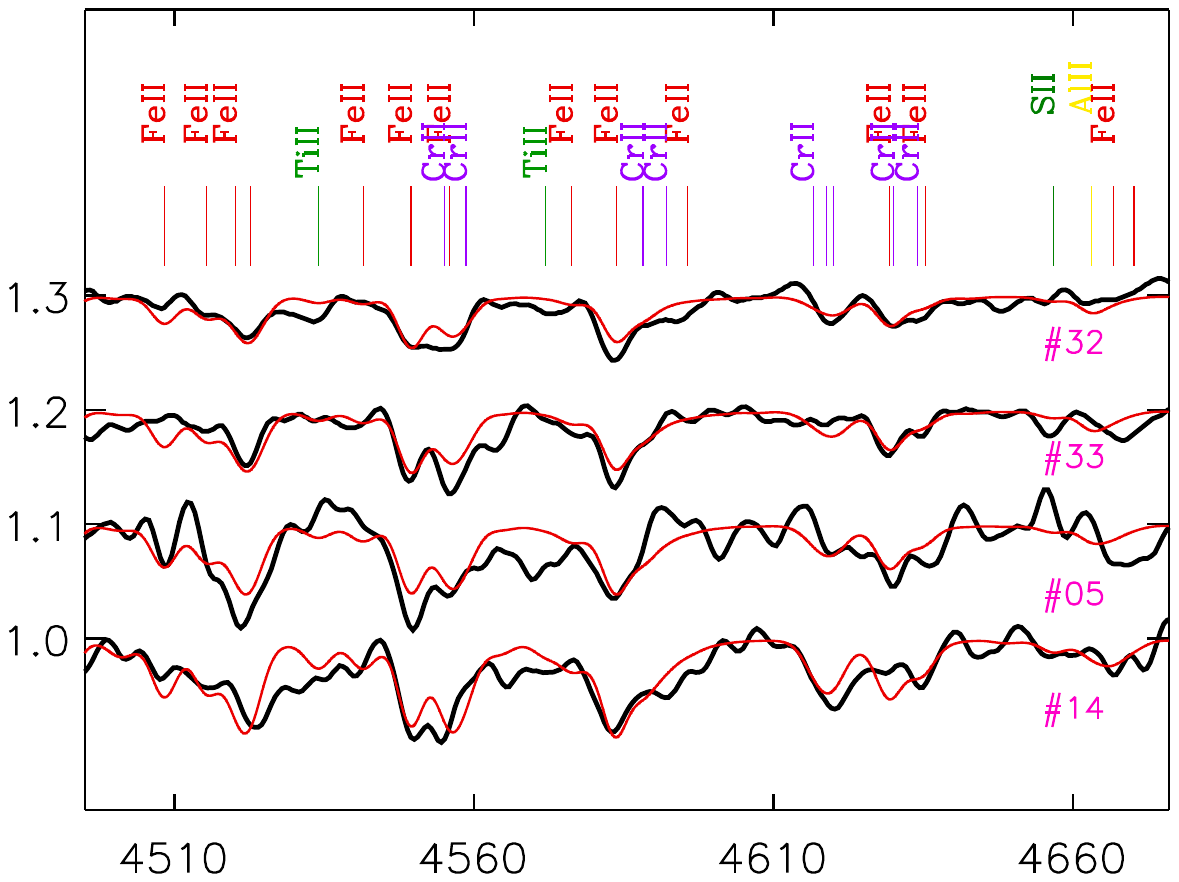}
    \includegraphics[width=0.445\textwidth]{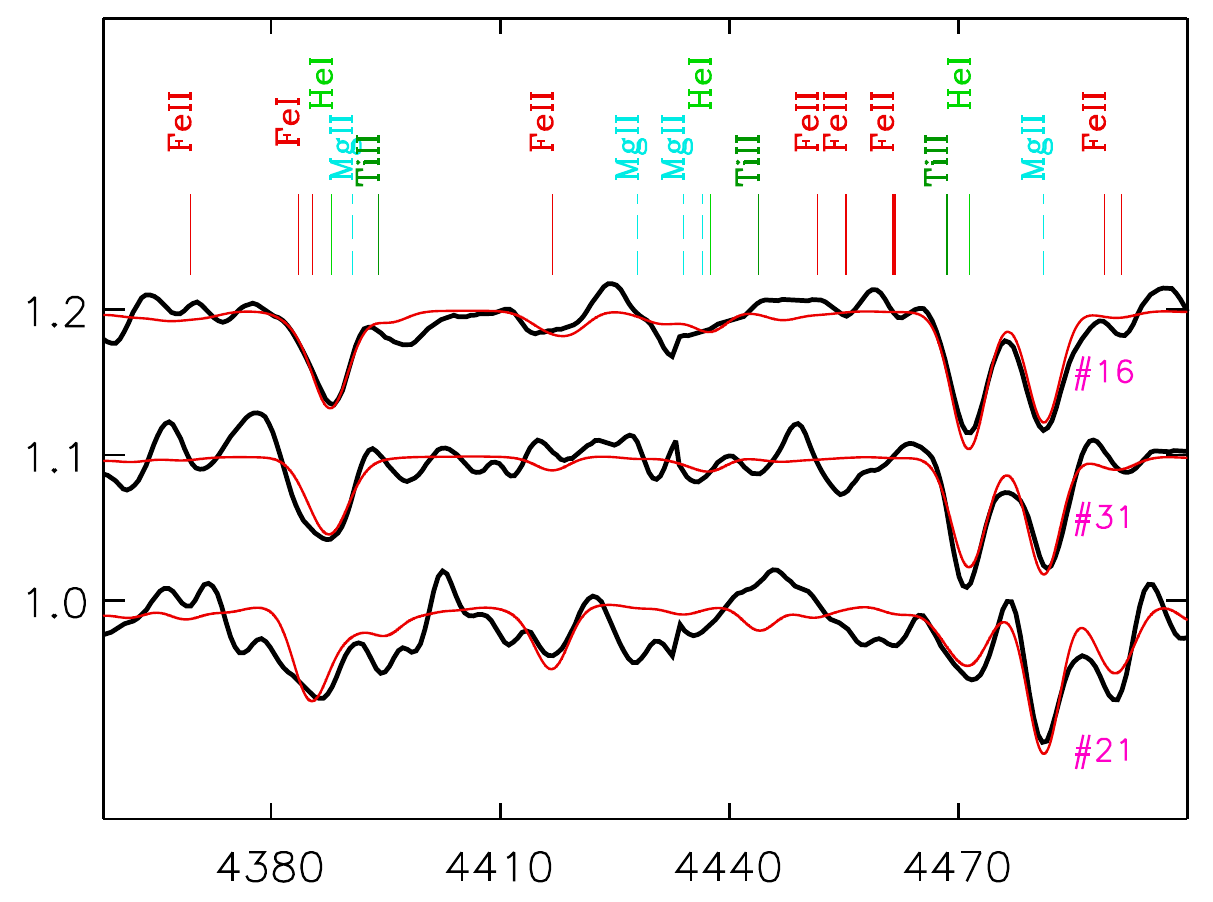}
    \caption{Examples of fits of the metal line spectrum. Upper panel: low temperature targets in the wavelength range from 4500 to 4670 \AA. Middle panel: same wavelength range but for targets with higher temperature. Note the different scale on the y-axis for the middle and upper panel. Bottom panel: two targets of high \teff\ (ID 16 and 31) and one target of medium \teff\ (ID 21)  in the range from 4360 to 4500 \AA. For better visibility the spectra are shifted vertically relative to each other by 0.2 in the upper panel and by 0.1 in the middle and lower panel. }
    \label{fig:metalfit}
\end{figure}

The results of the spectral analysis are given in Table~\ref{table:4}. An example of the spectral fits of the Balmer lines with the final model for three targets is shown in Figure~\ref{fig:Balmerfit}, while Figure~\ref{fig:metalfit} displays examples of the fits of the metal lines. Although NGC~4258 is the most distant galaxy studied so far in our project, the quality of the spectra and their corresponding fits are sufficient to derive stellar parameters with reasonable accuracy.

\begin{figure}[ht]
    \centering
    \includegraphics[width=0.47\textwidth]{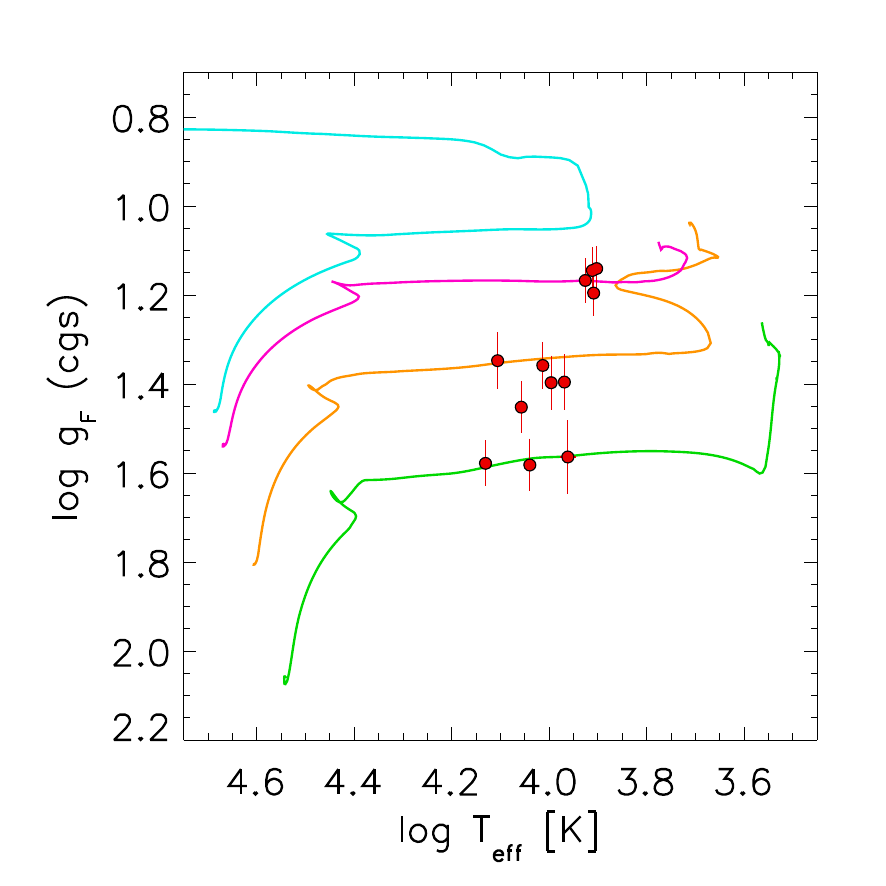}
    \caption{Spectroscopic Hertzsprung-Russell diagram (sHRD) of our targets in NGC~4258 together with MESA evolutionary tracks. The tracks have initial masses of 20 (green), 30 (orange), 50 (pink) and 60 (blue) $M_{\odot}$.} 
    \label{fig:sHRD}
\end{figure}

In addition to effective temperature, gravity and metallicity for each target Table~\ref{table:4} also contains flux-weighted gravities, defined as \loggf\,=\,\logg$-4\log(\teffmath/10^4)$. This quantity is crucial for the discussion of stellar evolution and distances. As found by \citet{Kudritzki:2003} and as shown in the papers mentioned above, \loggf\ can be used to determine accurate spectroscopic distances through the flux-weighted gravity--luminosity relation (\fglr) (see also \citealt{Kudritzki:2020}). We will determine an independent distance by this technique in Sec.~\ref{sec:FGLR}. In addition, the use of \loggf\ allows an investigation of the evolutionary status and stellar masses of our targets via the spectroscopic Hertzsprung-Russel diagram (sHRD), relating flux-weighted gravity and effective temperature, which has been introduced by \citet{Langer:2014}. While this diagram shows essentially the same behavior and information as the classical stellar HRD, it has the advantage that it is independent of the distance adopted. Our sHRD of NGC~4258 is shown in Figure~\ref{fig:sHRD} together with MESA evolutionary tracks (\citealt{Choi:2016}, \citealt{Dotter:2016}) of solar metallicity (see next section) including the effects of stellar rotation. The diagram indicates that our targets are massive stars of 20 \msun\ to 50 \msun\ having ages lower than 10 Myr. This is the same range of stellar masses and ages also encountered in our previous work. The only exception so far has been the extremely metal-poor dwarf galaxy Leo~A, where the masses turned out to be significantly lower (see \citealt{Urbaneja:2023} for a detailed discussion).

\begin{figure}[ht]
    \centering
    \includegraphics[width=0.47\textwidth]{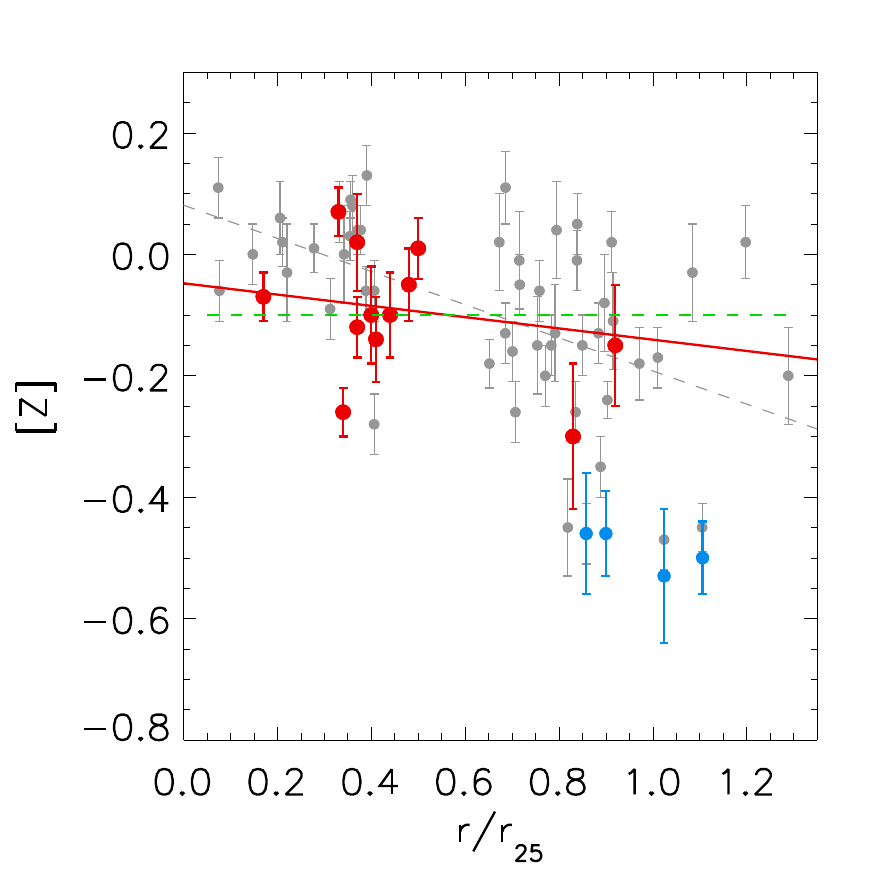}
    \ \par
    \ \par
    \includegraphics[width=0.47\textwidth]{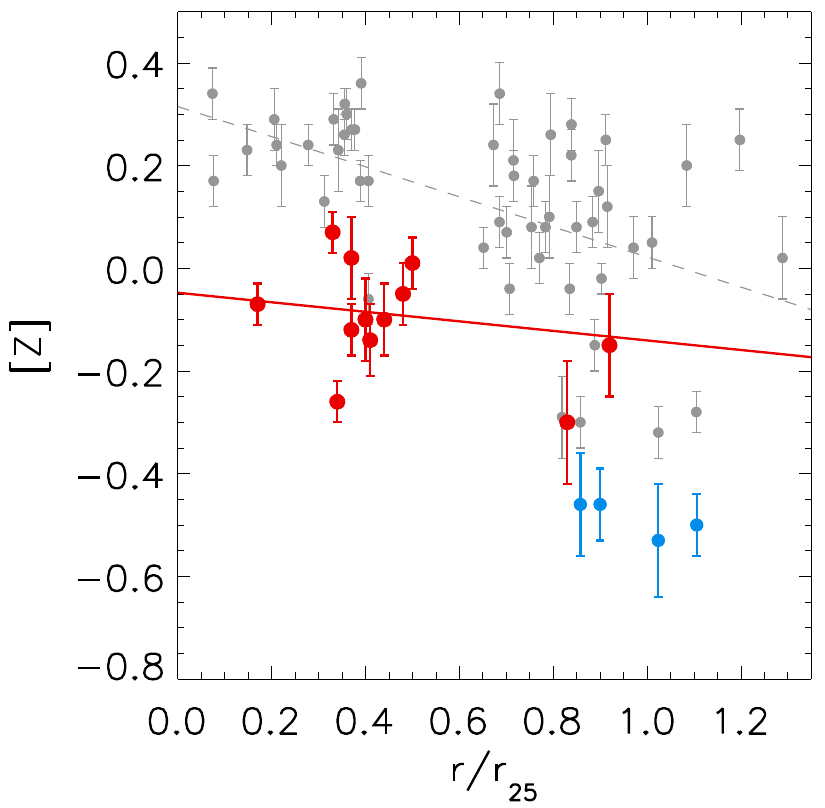}
    \caption{Blue supergiant metallicity $[Z]$ (red circles) as a function of deprojected galactocentric distance $r$/\rtf. The linear regression is plotted in red. \hii\ region oxygen abundances (in units of the solar value) by \citet{Yuan:2022} obtained from the strong line ratio \rtwothree~with the \citet{Pettini:2004} (upper plot) and \citet{Zaritsky:1994}  (lower plot) metallicity calibrations are shown in grey together with the corresponding regressions (dashed). The horizontal line in the upper plot indicates the Cepheid metallicity adopted by \citet{Riess:2022}. The blue symbols correspond to \hii\ region oxygen abundances obtained by the direct method using auroral lines \citep{Bresolin:2011}.} 
    \label{fig:metal1}
\end{figure}

Table~\ref{table:4} also contains bolometric corrections BC$_{F555W}$ calculated from the SEDs of the stellar atmosphere models for the final stellar parameters using the {\it HST} ACS F555W filter function. Once the interstellar reddening E({$\rm F435W-F555W$}) is determined (see Sec.~\ref{sec:FGLR}) the bolometric corrections are added to the dereddened F555W magnitudes to obtain the apparent bolometric magnitudes~\mbol~reported in the table. \mbol~and \loggf~will then be used in conjunction with the \fglr~to determine an independent distance to NGC~4258.

\section{Stellar metallicity, metallicity gradient and mass-metallicity relationship} \label{sec:metallicity}

The stellar metallicities $[Z]$ found for our targets vary from slightly super-solar down to sub-solar values. NGC~4258 is a relatively massive spiral galaxies with a stellar mass of $\log (M_*$/\msun) = 10.67 \citep{Leroy:2019} and, therefore, we expect a significant metallicity gradient with respect to the galactocentric distance of our targets. However, as Figure~\ref{fig:metal1} reveals, the metallicity gradient is very shallow. A linear regression yields $[Z] = -0.05\pm0.05 - (0.09\pm0.11)\,r$/\rtf. The gradient of $-0.09$ dex (\rtf)$^{-1}$ is significantly smaller than the average value of $-0.42$ dex (\rtf)$^{-1}$ found for massive spiral galaxies by, for instance, \citet{Ho:2015} from an investigation of a large sample of \hii\ regions in 49 galaxies. There are only very few galaxies in the Ho et al. sample that come close to this very low value. We also note that for all the other spirals studied so far in our blue supergiant spectroscopic work (NGC 300, NGC 55, NGC 3621, M33, M31, M83, MW) only NGC 55, with a gradient slope of $-0.22$ dex (\rtf)$^{-1}$, has a value comparably low within the error margins \citep{Kudritzki:2016}. The population synthesis study of the Great Barred Spiral NGC~1365 by \citet{Sextl:2024} yielded a gradient of $-0.59$ dex (\rtf)$^{-1}$ for the young stellar population. While it is beyond the scope of our present investigation, it will be worthwhile to study the disk growth and the star formation history of NGC~4258 in future work using the approach outlined by \citet{Kang:2021}, \citet{Kang:2023} or \citet{Sextl:2024}.

\begin{figure}[t]
    \centering
    \includegraphics[width=0.49\textwidth]{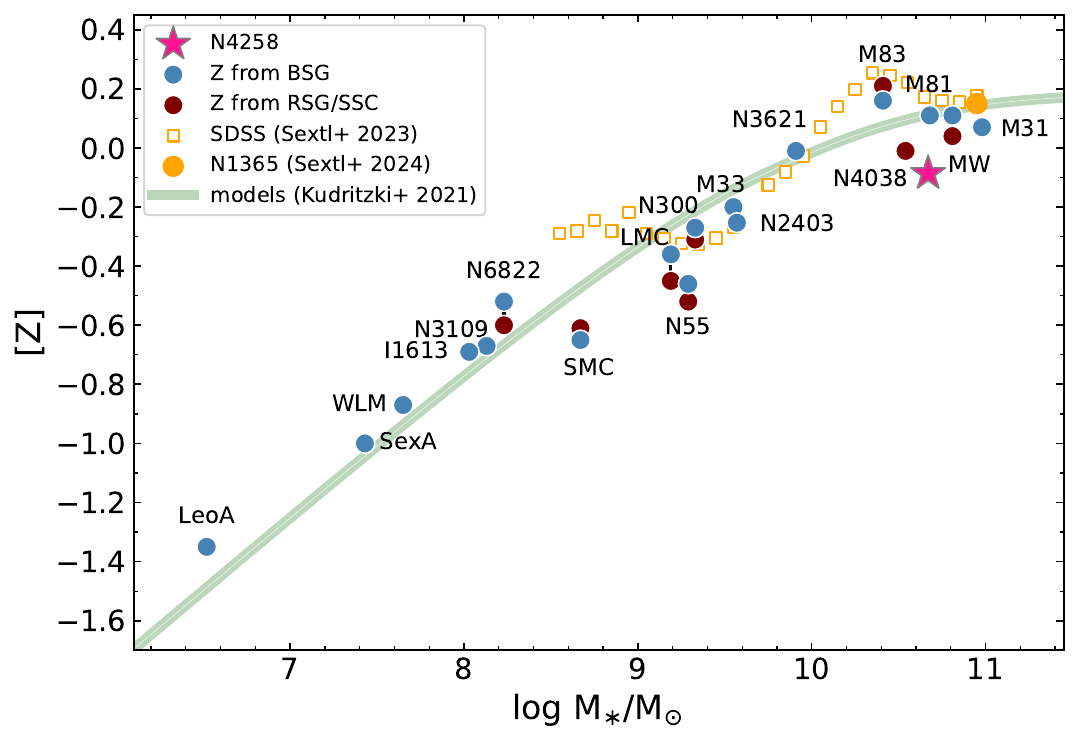}
    \caption{The mass-metallicity relationship of star-forming galaxies obtained from absorption line studies of the young stellar population. The pink star represents NGC~4258 as the result of this work. Blue circles refer to spectroscopy of blue supergiant stars (BSG) carried out in previous work as mentioned in the text. Red circles report metallicities determined by the analysis of NIR spectra of red supergiants (RSG) or superstar clusters (SSC). The yellow circle is the result of the spatially resolved stellar population synthesis study by \citet{Sextl:2024}, while the yellow squares refer to the analysis of integrated galaxy spectra of 250,000 star-forming SDSS galaxies again using a population synthesis technique \citep{sextl:2023}. Predictions from the galaxy evolution look-back models by \citet{Kudritzki:2021a,Kudritzki:2021b} are shown as the green line.} 
    \label{fig:MZR}
\end{figure}

Having encountered the unusual behaviour of the metallicity gradient, it is important to check how NGC~4258 behaves with respect to the mass-metallicity relationship (MZR) of star forming galaxies studied so far in our work. This is done in Figure~\ref{fig:MZR}. While located slightly below the other galaxies at comparable mass, NGC~4258 does not show any striking deviation from the observed MZR. As for all the other galaxies with a metallicity gradient in Figure~\ref{fig:MZR} we have used the regression metallicity at $r/$\rtf$=0.4$ for NGC~4258. For the original sources of the data displayed in Figure~\ref{fig:MZR} we refer to \citet{Urbaneja:2023} and \citet{Bresolin:2022} and references therein.

The metallicity of the young stellar population is important for distance determinations using Cepheid stars, because the zero point of their Period--Luminosity relation depends on metallicity (see, for instance, \citealt{Breuval:2021,Breuval:2024} and references therein). So far, in work beyond the Local Group the standard technique has been to infer metallicities from \hii\ region spectroscopy using strong emission lines and so-called strong line calibrations which calculate oxygen abundances from
observed emission line ratios. Unfortunately, as has been shown, for instance, by \citet{Kewley:2008} and \citet{Teimoorinia:2021}, the strong line calibrations have large systematic uncertainties, which in turn may affect Cepheid distance determinations. Stellar metallicities obtained from quantitative non-LTE spectroscopy have been carefully tested with a variety of techniques (see, for instance, \citealt{Gazak:2015}) and systematic effects are smaller than 0.1 dex in $[Z]$. For NGC~4258 as an anchor galaxy for the determination of the Hubble constant it is, thus, crucial to compare our stellar metallicities with the results obtained from \hii\ regions.

Comprehensive observations of \hii\ regions in NGC~4258 have been carried out by \citet{Bresolin:2011} and \citet{Yuan:2022} and oxygen abundances using different strong line calibrations have been provided in these works. Figures~\ref{fig:metal1} contains the \citet{Yuan:2022} oxygen abundances based on two strong line calibrations, those by \citet{Pettini:2004} and \citet{Zaritsky:1994}, respectively. Since the stellar metallicities are given in units of the solar metallicity we also plot the oxygen abundances in units of the solar value $12+\log$(N(O)/N(H))$_{\odot}$ = 8.69 \citep{Asplund:2009}.

The difference between these two calibrations is striking. While the \citet{Pettini:2004} calibration is in rough agreement with stellar work, the \citet{Zaritsky:1994} calibrated metallicities are larger by 0.2 to 0.3 dex. A similar result has been found by \citet{Bresolin:2016} in their detailed study of stellar and \hii\ region metallicities in M83.

The grey dashed lines in Figure~\ref{fig:metal1} correspond to the regression $[Z] = 0.08 - 0.27\,r/$\rtf\ given by \citet{Yuan:2022} for the \citet{Pettini:2004} calibrated oxygen abundances and to $[Z] = 0.31 - 0.29\,r/$\rtf\ for the \citet{Zaritsky:1994} calibration. These are slightly steeper gradients than found from our stellar blue supergiant analysis. The lower gradient obtained in the stellar work is very likely the reason for the very low metallicity dependence of the Cepheid PLR zero point found by 
\citet{Yuan:2022} when they compared Cepheid distance moduli as a function of galactocentric distance.

\begin{figure}[t]
    \centering
    \includegraphics[width=0.49\textwidth]{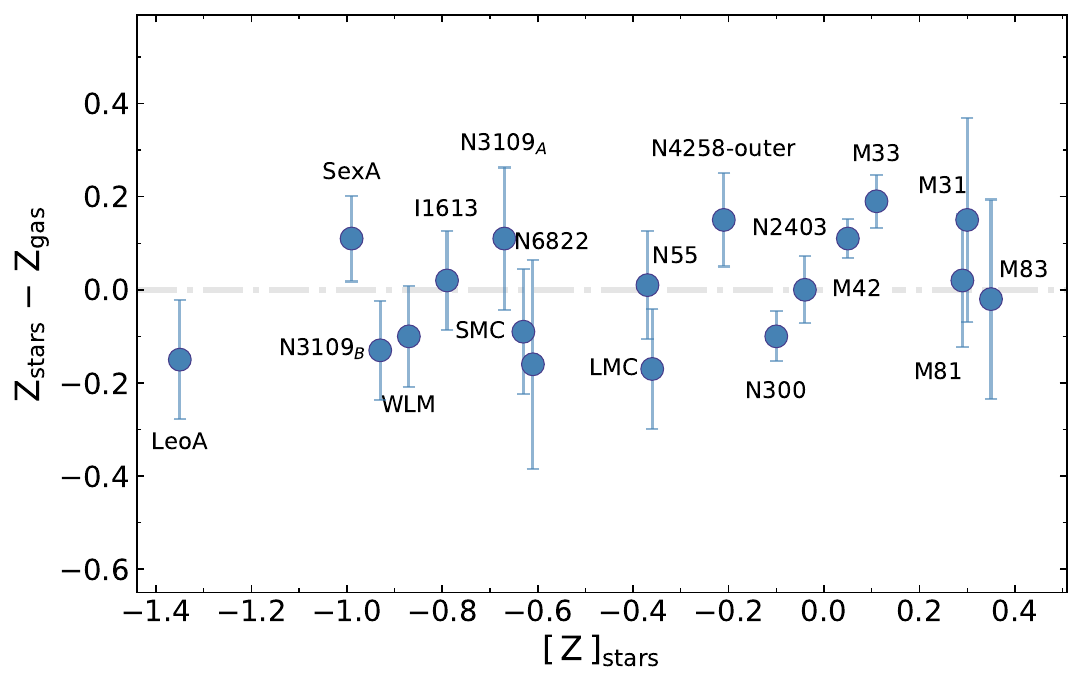}
    \caption{Difference between blue supergiant and \hii\ region metallicities. For the latter the direct method using collisionally excited auroral lines have been used. The \hii\ region nebular metallicities have been increased by 0.1 dex to account for oxygen depletion onto dust grains. The data are from \citet{Bresolin:2016,Bresolin:2022}, \citet{Urbaneja:2023} and this work.}
    \label{fig:Zau}
\end{figure}

Figure~\ref{fig:metal1} also shows metallicities of four outer \hii\ regions where a determination of the oxygen abundance by the more reliable direct method using auroral lines has been possible \citep{Bresolin:2011}. We quantify the difference between auroral line-based ionized gas metallicities and \bsg\ metallicities by considering the $Z$ offset between the two outermost \bsg s in Fig.~\ref{fig:metal1} and the two \hii\ regions located at similar galactocentric radii. 
It is well known that there is a small systematic difference of about 0.1 dex between stellar and direct method \hii\ region metallicities (see, for instance, \citealt{Bresolin:2016,Bresolin:2022}), which is attributed to the depletion of interstellar oxygen onto dust. The offset we find for NGC~4258, $0.25\pm0.10$ dex, seems somewhat large. On the other hand, as Figure~\ref{fig:Zau} reveals, there a few examples in our previous spectroscopic studies where a similar difference has been encountered. 
Note that in Figure~\ref{fig:Zau} the stellar metallicity employed for NGC~4258 refers to the measurement in the outermost region sampled by our observations. It is labelled as 'N4248-outer'.

As mentioned in the Introduction, NGC~4258 is a pivotal galaxy for the determination of the Hubble constant. The interferometrically resolved orbits of water masers in the central circumnuclear disk around the central black hole can be used to provide a geometric determination of the angular-diameter distance to the galaxy with a precision of 1.5 percent \citep{Reid:2019}. In this way, NGC~4258 has become an anchor point for the calibration of stellar distance indicators such as Cepheids \citep{Riess:2022} and red giant stars (\citealt{Jang:2021,Anand:2024}). In this regard, the metallicity of the young stellar population of NGC~4258 is crucially important, because the zero point of the Cepheid Period--Luminosity Relation (PLR) depends on metallicity. The calibrated PLR zero point metallicity dependence is then used for distance determinations.

\citet{Riess:2022}, in their comprehensive study on the measurement of the Hubble constant, have used three galaxies as Cepheid PLR anchor points: the Milky Way, the LMC, and the maser galaxy NGC~4258. While they could use stellar metallicities for the former two, they had to rely on \hii\ region metallicities as a stellar proxy for the latter. In view of the uncertainties of the individual nebular strong line calibrations, they did not rely on a single calibration but instead used an average of nine calibrations as provided by \citet{Teimoorinia:2021}, which yielded a stellar metallicity of $[Z] = -0.10$ (see Table 3 in \citealt{Riess:2022}) based on \hii\ region emission line flux measurements by \citet{Yuan:2022}. As pointed out in their paper, this calibration gives metallicities very similar to the \citet{Pettini:2004} calibration.

Our direct determination of stellar metallicities from blue supergiant stars provides important new input for this calibration process. First, the very shallow stellar $[Z]$ gradient justifies the use of one single metallicity value for NGC~4258. Second, the value of $[Z] = -0.1$ is in good agreement with an average over the spatial regions where Cepheids are encountered. It puts the $[Z]$ calibration of the PLR zero point on a purely stellar basis and confirms the \citet{Riess:2022} result.

When applying the $[Z]$ dependence of the PLR zero points to their distance determination of galaxies with Cepheids, \citet{Riess:2022} use again \hii\ region emission line fluxes using the same calibration as described above. This is a revision from previous work \citep{Riess:2016}, where the \citet{Zaritsky:1994} calibration has been used. As is very obvious from Figure~\ref{fig:metal1} and our previous work, this calibration leads to metallicities which are systematically too high. The revision to the new calibration, on the other hand, brings the estimated metallicities much closer to the values expected from stellar spectroscopy. 


\section{Spectroscopic distance using the FGLR method} \label{sec:FGLR}

Blue supergiant stars follow a relationship (\fglr) between flux-weighted gravity \loggf\ as defined in Sec.~\ref{sec:quantitative} and absolute bolometric magnitude \citep{Kudritzki:2003, Kudritzki:2008}. With the calibration by \citet{Urbaneja:2017} based on a spectroscopic study of 90 BSGs in the LMC and the new precision distance to the LMC by \citet{Pietrzynski:2019} the \fglr~is given by
\begin{equation}
    \log\,g_F\, \geq 1.30:\,   M_{\rm bol}\,= 3.20(\log\,g_F\,-\,1.3)\,- 8.518
\end{equation}
\begin{equation}
    \log\,g_F\, \leq 1.30:\,   M_{\rm bol}\,= 8.34(\log\,g_F\,-\,1.3)\,- 8.518
\end{equation}
and has been successfully used for spectroscopic distance determinations out to 6 Mpc so far (see \citealt{Bresolin:2022} for a summary). 
The application of the \fglr~to determine a spectroscopic distance to NGC~4258 with its precise geometrical distance is a crucial test of the method.

In a first step, we need to determine interstellar reddening and extinction. For this purpose we use the observed {\it HST} and {\it JWST} photometry described in Sec.~\ref{sec:observations}. We compare with synthetic {\it HST} and {\it JWST} colors obtained from reddened spectral energy distributions which we calculated with our model atmospheres for the final stellar parameters \teff, \logg~and $[Z]$ given in Table~\ref{table:4}. We apply four different monochromatic reddening laws \citep{Cardelli:1989, Fitzpatrick:1999,Maiz:2014, Maiz:2017, Odonnell:1994} and calculate reddened colors ($\rm F435W-F555W$), ($\rm F555W-F814W$), ($\rm F555W-F160W$), and if available ($\rm F555W-F070W$), ($\rm F555W-F090W$), ($\rm F555W-F115W$), ($\rm F555W-F150W$), ($\rm F555W-F227W$), ($\rm F555W-F356W$)  for a grid of monochromatic reddening values E($\rm 4405-5495$) and adopted ratios of monochromatic extinction to reddening of R$_{\rm 5495}$ = A$_{5495}$/E($\rm 4405-5495$). A $\chi^2$ comparison between observed and calculated colors provides the best values of monochromatic reddening and R$_{\rm 5495}$, which are then used to calculate the filter pass-band reddening E($\rm F435W-F555W$) together with extinction A$_{\rm F555W}$ and the ratio R$_{\rm F555W}$ = A$_{\rm F555W}$/E($\rm F435W-F555W$). In this procedure, we deliberately allow for R$_{\rm 5495}$ to deviate from the commonly used "standard" value of 3.1. This is motivated by the spectroscopic studies by \citet{Urbaneja:2017,Maiz:2017,sextl:2023,Sextl:2024} which show that in star forming galaxies a wide range of the ratio of reddening to extinction is encountered. Thus, E($\rm 4405-5495$) and R$_{\rm 5495}$ are determined simultaneously. The approach is the same as described in \citet{Urbaneja:2017} and \citet{Taormina:2020,Taormina:2024}. We note that the determination of R$_{\rm 5495}$ is only possible because we have photometry available at 1.6 $\mu$m and for some targets at even longer wavelengths. While the errors of R$_{\rm F555W}$ are relatively large, the extinction A$_{\rm F555W}$ remains well constrained because of the anticorrelation of E($\rm F435W-F555W$) and R$_{\rm F555W}$ in each $\chi^2$ fit (see \citealt{Urbaneja:2017}, their Figure 9 and Sec.~4.4).

\begin{figure}[t]
    \centering
    \includegraphics[width=0.45\textwidth]{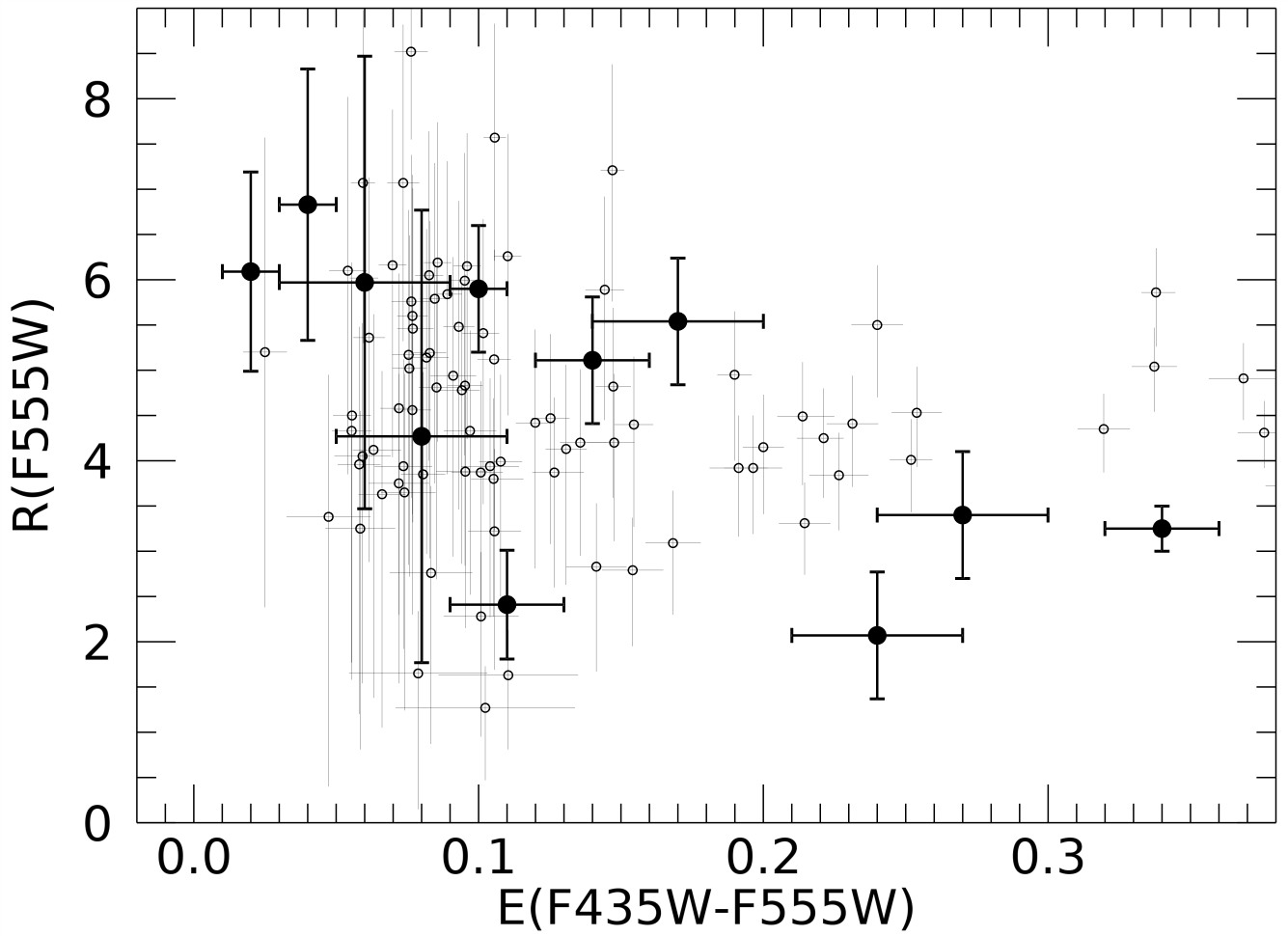}
    \caption{The observed distribution of $R_{F555W}$ values as a function of $E_{F435W-F555W}$ in NGC~4258, plotted using filled symbols. For reference, the values derived in the LMC by \citet{Urbaneja:2017} are plotted using smaller open symbols.}
    \label{fig:EvR}
\end{figure}

The reddening values resulting from our fit are also included in Table~\ref{table:4} and plotted in Figure~\ref{fig:EvR}. They are obtained with the O'Donnell reddening law, but the other three reddening laws give very similar results.
In the next step, we apply bolometric corrections BC$_{F555W}$ (see Sec.~\ref{sec:quantitative} and Table \ref{table:4}) to the dereddened $m_{F555W}$ magnitudes of our BSG targets to obtain apparent bolometric magnitudes \mbol. The latter are also given in Table~\ref{table:4}. As a safety check we repeat the determination of \mbol~replacing the F555W magnitudes by F606W for the seven targets in Table~\ref{table:2}, which have measured magnitudes in this passband and which are used for the \fglr~fit. The resulting \mbol~are very similar with the average difference between the two approaches smaller than 0.01 magnitude. 

\begin{figure}[t]
    \centering
    \includegraphics[width=0.45\textwidth]{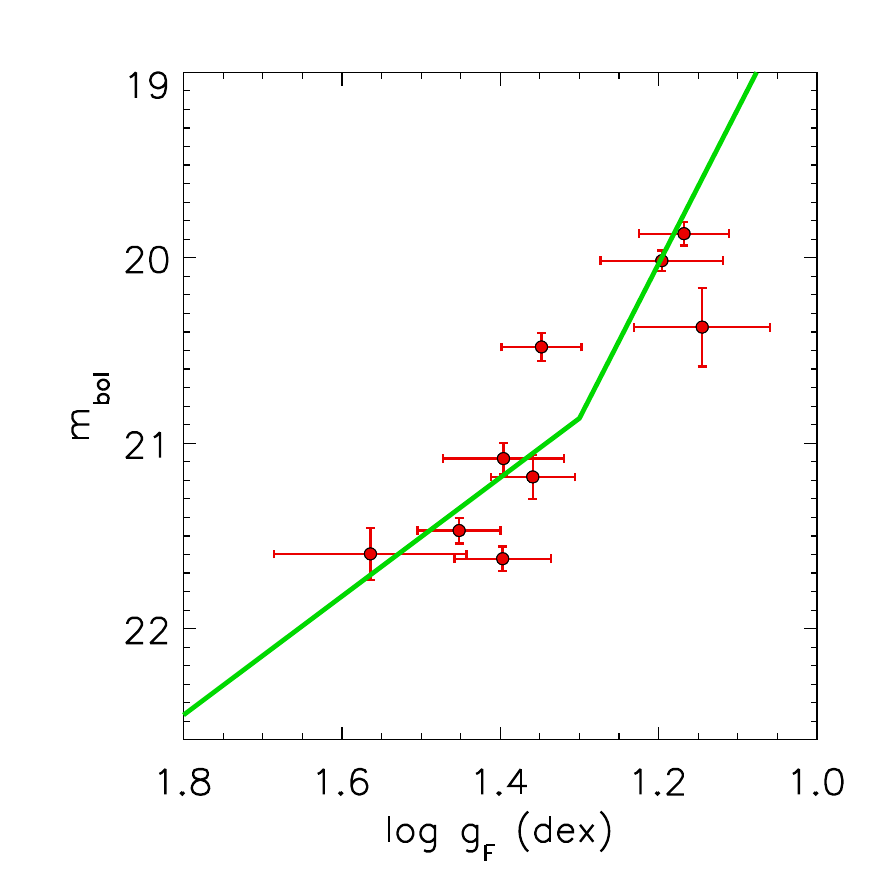}
    \caption{The observed \fglr\ in NGC~4258. The fiducial relationships of Equations (1) and (2) shifted by the fitted distance modulus are shown in green. A distance modulus of $m-M = 29.38\pm0.12$ is obtained from the fit.}
    \label{fig:FGLR}
\end{figure}

In the final step, we then fit the fiducial \fglr~of equations (1) and (2) to the data to determine a distance modulus. We check for $\ge$ 2$\sigma$ magnitude outliers and identify ID 8 and 32 as too bright (unresolved stellar multiplicity) and ID 26 as too faint (possibly the result of strong mass loss or binary evolution). After removal of these targets the \fglr~fit yields a distance modulus of $m-M = 29.38\pm0.12$, in good agreement with the VLBI geometrical distance. The resulting observed \fglr~ in apparent bolometric magnitude is given in Figure~\ref{fig:FGLR} together with the fiducial fits of equations (1) and (2).

\section{Summary and discussion} \label{sec:summary}

With the results obtained in the previous sections we have been able to characterize the properties of the young stellar population in the Hubble constant anchor point galaxy NGC~4258. The blue supergiants investigated are less than 10 Myr old and have masses in the range of 20-50 \msun. The metallicity of these young massive stars is slightly less than solar, in agreement with the mass-metallicity relationship of star forming galaxies obtained from absorption line studies of the young stellar population. We find a very shallow metallicity gradient, unusual for massive spiral galaxies. A follow-up investigation of the history of star formation and stellar disk growth applying population synthesis techniques (see, for instance, \citealt{Sextl:2024}) and a galaxy evolution modelling approach \citep{Kang:2023} appears to be worthwhile.

A detailed comparison with \hii\ region emission line studies allows to assess the systematic uncertainties of the strong line abundance determination methods employed and confirms that the calibration by \citet{Pettini:2004} in the new version by \citet{Teimoorinia:2021} is in reasonable agreement with stellar metallicities. Most importantly, our results put the calibration of the metallicity of the Cepheid Period--Luminosity relation in this crucial anchor point galaxy on a purely stellar basis and confirm the values adopted by \citet{Riess:2022} in their measurement of the Hubble constant.

We use stellar temperatures, gravities and dereddened apparent bolometric magnitudes to determine an independent distance to NGC~4258 using the blue supergiant \fglr~method. The value obtained is in good agreement with the very accurate geometrical distance based on {\small VLBI} water maser interferometry. It also agrees with the \trgb\ distance obtained by \citet{Anand:2021}. This confirms the reliability of the \fglr~method as an independent distance indicator. As a spectroscopic technique, which allows for a reliable direct measurement of reddening, extinction and metallicity, the \fglr~method will provide an important alternative for extragalactic distance determinations in the forthcoming era of extremely large telescopes.

\ \par
This research has made use of the Keck Observatory Archive (KOA), which is operated by the W. M. Keck Observatory and the NASA Exoplanet Science Institute (NExScI), under contract with the National Aeronautics and Space Administration.
RPK acknowledges support by the Munich Excellence Cluster Origins and the Munich Institute for Astro-, Particle and BioPhysics (MIAPbP) funded by the Deutsche Forschungsgemeinschaft (DFG, German Research Foundation) under Germany's Excellence Strategy EXC-2094 390783311.
The authors wish to recognize and acknowledge the very significant cultural role and reverence that the summit of Maunakea has always had within the indigenous Hawaiian community. We are most fortunate to have the opportunity to conduct observations from this mountain.

\facility{Keck:I (LRIS)}

\software{IRAF (\citealt{Tody:1986, Tody:1993, Fitzpatrick:2024}), DAOPHOT \citep{Stetson:1987}, ALLSTAR \citep{Stetson:1994}, SciPy (\citealt{Virtanen:2020}), NumPy (\citealt{Harris:2020}), Matplotlib (\citealt{Hunter:2007a}), PyRAF (\citealt{Science-Software-Branch-at-STScI:2012}), DrizzlePac \citep{DrizzlePac}, TinyTim \citep{Krist:1993}, DOLPHOT \citep{Dolphin:2000,Dolphin:2016,Weisz:2024}}

\bibliographystyle{aasjournal}
\bibliography{Papers.bib}

\end{document}